\documentclass[aps,prb,preprint,showpacs,groupedaddress]{revtex4}
\usepackage[dvips]{graphicx}
\begin{document}
\title{Contrasting the magnetic response between magnetic-glass and reentrant spin-glass}
\author{S. B. Roy}
\author{M. K. Chattopadhyay}
\author{}
\affiliation{Magnetic and Superconducting Materials Section, Raja Ramanna Centre for 
Advanced Technology, Indore 452013, India.}
\date{\today}
\begin{abstract}
Magnetic-glass is a recently identified phenomenon in various classes of magnetic systems undergoing a first order magnetic phase transition. We shall highlight here a few experimentally determined characteristics of magnetic-glass and the relevant set of experiments, which will enable to distinguish a magnetic-glass unequivocally from the well known phenomena of spin-glass and reentrant spin-glass.   
\end{abstract} 
\maketitle
It has been shown recently that in many magnetic systems a kinetic arrest of the first order ferromagnetic (FM) to antiferromagnetic (AFM) phase transition leads to a non-equilibrium magnetic state with a configuration of FM and AFM clusters frozen randomly in experimental time scale \cite{1,2,3,4,5,6,7}. The dynamics of this non-equilibrium magnetic state is very similar to that of a structural glass \cite{8}, and analogically this new magnetic state is named magnetic-glass \cite{2,7,9}. The results emerging from disparate classes of magnetic systems starting from alloys and intermetallic compounds \cite{1,2,3,6,7,9} to manganite systems showing colossal magnetoresistance (CMR)\cite{3,4,5} suggest that this magnetic-glass phenomenon is independent of the underlying microscopic nature of magnetic interactions. Analogous to the structural glasses, the magnetic-glass (MG) can undergo devitrification with the change in temperature (T)\cite{10,11}.  

Competition between AFM and FM interactions plays the central role in spin-glass (SG) and reentrant spin-glass (RSG) \cite{12,13}. In SG this competition is so strong that none of the long range magnetic orders is established, instead it gives rise to a random spin configuration frozen in time \cite{13}. In RSG long range magnetic order (FM or AFM) appears in certain T regime. However, the competing interactions introduce some frustration amongst the set of spins, which ultimately leads to the partial or total breakdown of the higher T FM or AFM state to a SG like state at the lowest T \cite{12,13}.  The spin-configuration of this lower T RSG state consists of individual spins (or small spin-clusters) frozen randomly in the microscopic scale with \cite{14} or without \cite{15} a trace of long-range FM order along the direction of the applied magnetic field (H).

The onset of both of these non-trivial MG and RSG state is accompanied by distinct H and T history dependence of bulk magnetic response i.e. thermomagnetic irreversibilities (TMI) and metastability, which at first sight can appear to be quite similar in nature. Such TMI and metastability are very well studied experimental observables in SG and RSG systems \cite{12,13}, and they are regularly used for initial identification of SG and RSG behaviour in a new magnetic system. The main aim of the present work is to carefully study and compare the TMI and metastability associated with the MG and RSG behaviour. We shall then highlight the identifiable features in such experimental observables, which will enable to distinguish a MG unequivocally from RSG. 

For our comparative study we have chosen a well studied MG system Ce(Fe$_{0.96}$Ru$_{0.04}$)$_2$  \cite{2,16} and a canonical RSG system Au$_{82}$Fe$_{18}$ alloy \cite{12,13}.  The FM-RSG transition in AuFe alloys above the percolation concentration of 15\% Fe has been studied in great details through both bulk properties and microscopic measurements \cite{17}. Various theoretical models have been proposed to understand these experimental results \cite{14,15}.  In Ce(Fe$_{0.96}$Ru$_{0.04}$)$_2$ the low T state is AFM in zero and relatively low ($\leq$ 10 kOe) applied H \cite{2,16}. In the presence of an applied field H$>$ 10 kOe the first order FM-AFM transition in Ce(Fe$_{0.96}$Ru$_{0.04}$)$_2$ gets kinetically arrested giving rise to a MG state \cite{2}. We shall now present below the contrasting TMI and metastablities associated with the RSG behaviour in Au$_{82}$Fe$_{18}$, and MG behaviour in Ce(Fe$_{0.96}$Ru$_{0.04}$)$_2$.

\begin{figure}[t]
\centering
\includegraphics[width = 8 cm]{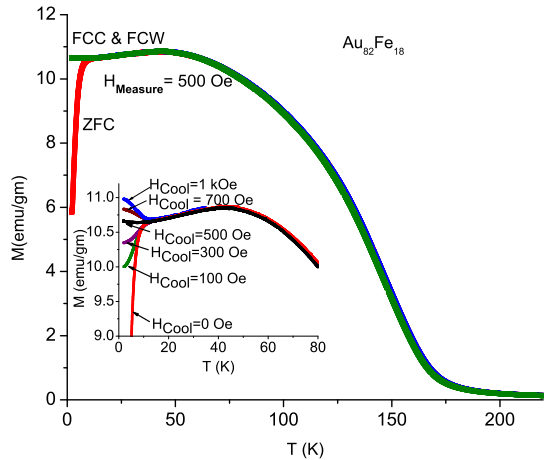}
\caption{(Color online):M vs T plot for Au$_{82}$Fe$_{18}$ obtained with ZFC, FCC and FCW protocol with H=500 Oe. The inset shows M vs T plot obtained under an experimental protocol of 'cooling and heating in unequal field' with H$_{Measure}$=500 Oe. See text for details.}
\label{fig1}
\end{figure}
The details of the preparation and characterization of the Ce(Fe$_{0.96}$Ru$_{0.04}$)$_2$ and Au$_{82}$Fe$_{18}$ samples used here can be found in references \cite{16} and \cite{18} respectively. The Au$_{82}$Fe$_{18}$ sample, however, was freshly annealed at 800$^0$C for 6 hours and quenched in liquid nitrogen before starting the present experimental cycle. Bulk magnetization measurements were made with a commercial vibrating sample magnetometer (VSM;Quantum Design, USA). We use three experimental protocols, zero field cooled (ZFC), field cooled cooling (FCC) and field cooled warming (FCW), for magnetization (M) measurements. In the ZFC mode the sample is cooled to the lowest T of measurement before the applied H is switched on, and the measurement is made while warming up the sample. In the FCC mode the applied H is switched on in the T regime above the FM-AFM transition temperature in the case of Ce(Fe$_{0.96}$Ru$_{0.04}$)$_2$ and FM-RSG transition temperature in the case of Au$_{82}$Fe$_{18}$, and the measurement is made while cooling across the transition temperature to the lowest T of measurement. After completion of measurement in the FCC mode, the data points are taken again in the presence of same applied H while warming up the sample. This is called FCW mode. A fixed rate of T variation 1K/Min has been used all throughout the present study.

\begin{figure}[t]
\centering
\includegraphics[width = 8 cm]{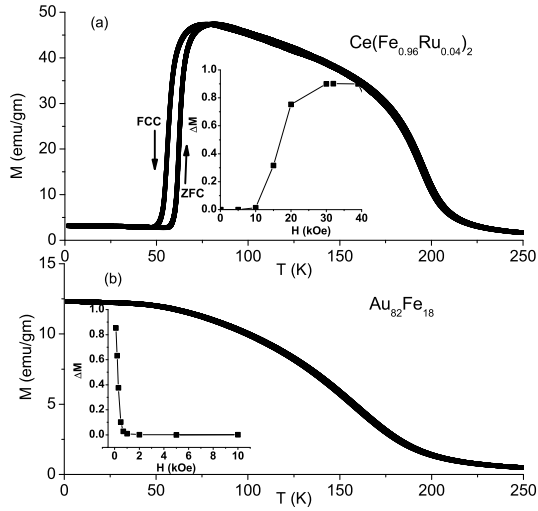}
\caption{M vs T plots for (a) Ce(Fe$_{0.96}$Ru$_{0.04}$)$_2$ (b) Au$_{82}$Fe$_{18}$  obtained with H= 5 kOe. Insets of Fig. 2(a) ( 2(b) ) show the difference $\Delta$M between M$_{FCC}$(T) and M$_{ZFC}$(T) (normalized with respect to M$_{FCC}$(T))  as a function of H for Ce(Fe$_{0.96}$Ru$_{0.04}$)$_2$ (Au$_{82}$Fe$_{18}$).} 
\label{fig2}
\end{figure}
The main frame of Fig. 1 presents the M versus T plot of Au$_{82}$Fe$_{18}$ alloy in H=500 Oe, obtained under the ZFC, FCC and FCW mode. The value of Curie temperature (T$_C$ $\approx$155K) estimated from the point of inflection in the M-T curve matches well with the earlier reported value in the literature \cite{17}. The onset of the FM-RSG transition is marked by a small but distinct maximum in the M-T curve at a temperature T$_M$ $\approx$50K. Then at a further lower temperature (T$_{irrv}$) there is a sharp drop in the ZFC M-T curve accompanied by a clear bifurcation of the ZFC and FC M(T) curves. This maximum in M(T) and the onset of strong TMI at T$_{irrv}$ are the hallmarks of RSG behaviour \cite{12}. Both these features are explained within a mean field theory of second order phase transition \cite{14} . There is also an alternative viewpoint, where the maximum at T$_M$ is envisaged as due to the onset of random freezing of isolated Fe-clusters in Au$_{82}$Fe$_{18}$, which in turn creates a random internal field acting on the infinite FM cluster and leading to a complete breakdown of the long range FM order into a spin-glass state at a lower T \cite{15}.  Note that in the main frame of Fig. 1 the M-T curves obtained under the FCC and FCW protocol completely overlap, and this is in consonance with both the types of theoretical pictures \cite{14,15}. With the increase in H, T$_{irrv}$ decreases and the M-T curve with H= 5 kOe (see the mainframe of Fig. 2(b)) resembles that of a standard FM with no trace of FM-RSG transition at least down to 2K. 

The main frame of Fig. 2(a) presents the M versus T plot of Ce(Fe$_{0.96}$Ru$_{0.04}$)$_2$ in H=5 kOe, obtained under the ZFC, FCC and FCW mode. Note that here the M$_{ZFC}$ (T) merges with M$_{FCW}$(T) at all T of measurement. A sharp rise (fall) in M in ZFC (FCC) path (see Fig. 2(a)) at temperatures T$_{NW}$(T$_{NC}$) around 65K marks the onset of AFM-FM (FM-AFM) transition while warming (cooling) \cite{16}. The distinct thermal hysteresis between M$_{FCC}$(T)and M$_{ZFC}$(T)(or M$_{FCW}$(T)) in the transition region arises due to the first order nature of the FM-AFM phase transition in Ce(Fe$_{0.96}$Ru$_{0.04}$)$_2$. The end point of the thermal hysteresis while cooling (warming) represents the limit of supercooling T* (superheating T**) across the first order phase transition \cite{16}.  Below (above)  T* (T**) the system is in the equilibrium AFM (FM) state.  

In the FCC mode above a critical applied H of 10 kOe, the FM-AFM transition gets kinetically arrested leading to the formation of the MG state \cite{2}. This behaviour is shown in the mainframe of Fig.3 in the M vs T plot of Ce(Fe$_{0.96}$Ru$_{0.04}$)$_2$ in a field of 20 kOe. The conversion to low T AFM state is not completed in the FCC mode. While warming up part devitrification of the MG state (to equilibrium AFM state) occurs and the system eventually reaches back to the higher T FM state. In contrast, in the ZFC mode the applied H is switched on at the lowest T of measurement, and since in H$\leq$10 kOe there is no formation of MG, the equilibrium AFM state can be reached and subsequently transformed with the increase in T to the FM state. All these effects give rise to interesting TMI where M$_{FCC}$(T)$\neq$ M$_{FCW}$(T) (and M$_{ZFC}$(T)) over a large T regime (see mainframe of Fig. 3). This onset of the MG state in an applied H, can be compared with the recent observation of the formation of glassy sate in liquid Ge under external pressure \cite{19}. 

In striking contrast to the TMI in the RSG state of Au$_{82}$Fe$_{18}$, the TMI associated with the MG behaviour in Ce(Fe$_{0.96}$Ru$_{0.04}$)$_2$ appears only above a certain critical H, and its magnitude increases with H. To highlight this difference in TMI we plot in the inset of Fig.2(a) and 2(b) $\Delta$M = (M$_{FCC}$(T)- M$_{ZFC}$(T))$/$M$_{FCC}$(T) measured at 5 K, as a function of applied H both for Ce(Fe$_{0.96}$Ru$_{0.04}$)$_2$ and Au$_{82}$Fe$_{18}$. In Au$_{82}$Fe$_{18}$  $\Delta$M falls to zero rapidly as H increase to 5 kOe, while in Ce(Fe$_{0.96}$Ru$_{0.04}$)$_2$ $\Delta$M acquires non-zero value only above H=10 kOe, and increases thereafter with the further increase in H.

\begin{figure}[t]
\centering
\includegraphics[width = 8 cm]{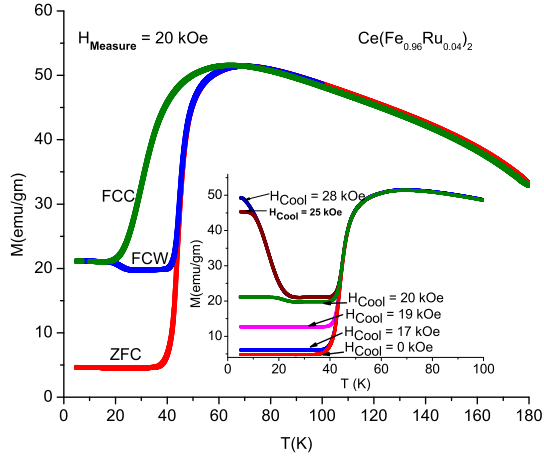}
\caption{(Color online):M vs T plot for Ce(Fe$_{0.96}$Ru$_{0.04}$)$_2$  obtained with ZFC, FCC and FCW protocol with H=20 kOe. The inset shows M vs T plot obtained under an experimenatl protocol of 'CHUF' with H$_{Measure}$=20 kOe. See text for details.}
\label{fig3}
\end{figure}
The quenched disorder in the concerned magnetic systems influences the FM-AFM first order transition process, and introduces a landscape of transition temperature T$_N$ \cite{20}. In such systems the H-T phase diagram consists of the bands of transition temperature (T$_N$), supercooling and superheating  limit (T* and T**) and a kinetic arrest temperature band (T$_K$) below which the system enters a MG state \cite{1,3,10,21}. The correlation between the characteristic temperatures T$_N$, T* (T**) and T$_K$ and its experimental consequences have been studied with a newly introduced experimental protocol, where the system is cooled across the transition temperature in certain applied H$_{Cool}$ and the magnetization studies are made while warming and after changing this H$_{Cool}$ isothermally to a different H$_{Measure}$ ( higher or lower than H$_{Cool}$) at the lowest T of measurement \cite{22}. This experimental protocol is in contrast with the standard field cooling protocols FCC and FCW, where the H$_{Cool}$ and the H$_{measure}$ while warming is the same. This technique of 'cooling and heating in unequal field (CHUF)' has been used to investigate the MG phenomenon in various CMR-manganite systems \cite{22}. It has been shown clearly that in a kinetically arrested FM-AFM transition, while warming with H$_{Measure} >$ H$_{Cool}$(H$_{Measure}<$H$_{Cool}$) under the CHUF protocol, one observes only one sharp structure (two sharp structures) in M(T)\cite{22}. We use this key result here to discern between a MG and RSG. The inset of Fig. 3 shows the results of M(T) measurements in Ce(Fe$_{0.96}$Ru$_{0.04}$)$_2$ obtained under CHUF protocol with H$_{Measure}$=20 kOe. With H$_{Measure}>$ H$_{Cool}$, there is only one sharp rise in M(T) leading to the FM state. On the other hand, when H$_{Measure}<$ H$_{Cool}$ the M(T) drops sharply and flattens before rising sharply again to reach the FM state. With a higher value of H$_{Cool}$ the state prepared at the lowest T has a large fraction of the  FM component in the MG state, and in the FCW mode M(T) drops sharply with rising T due to the devitrification of this non-equilibrium FM component. For a more detailed explanation of the origin of such distinct characteristic features associated with the MG phenomenon, the reader is referred to the Ref.22.The change in sign of the inequality between H$_{Measure}$ and H$_{Cool}$ does not lead to such distinctive characteristic features in the RSG system Au$_{82}$Fe$_{18}$, and this is shown experimentally in the M(T) study with H$_{Measure}$=500 Oe (see the inset of Fig.1). The algebraic value of TMI around T$_{irrv}$ changes monotonically with the change in sign of the inequality.

\begin{figure}[t]
\centering
\includegraphics[width = 8 cm]{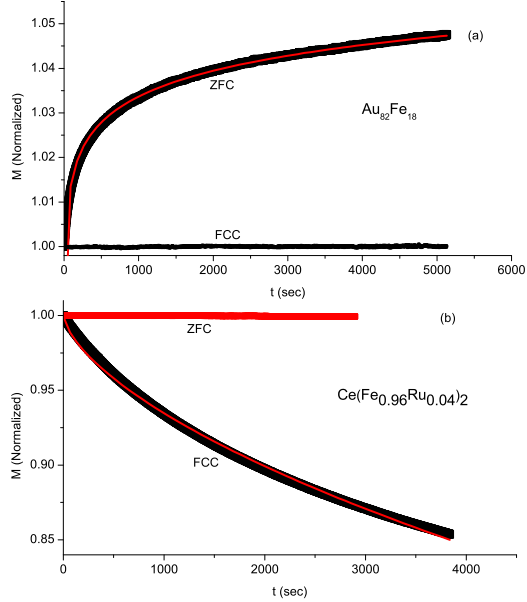}
\caption{(Color online):M vs time plots for (a)Au$_{82}$Fe$_{18}$ at T=14K (b)Ce(Fe$_{0.96}$Ru$_{0.04}$)$_2$ at T=18K  obtained under ZFC and FCC protocols. M is normalized with respect to the initial M$_0$ value obtained after 1 second of stabilizing at the respective T of measurement, which is reached with a cooling rate of 1K/min in the 'temperature no-overshoot' mode of the instrument. The relaxation data for Au$_{82}$Fe$_{18}$ (Ce(Fe$_{0.96}$Ru$_{0.04}$)$_2$) are fitted with the eequation M(t)= -1 +2t$^\gamma$ (stretched exponential function M(t) $\propto$ exp[-(t$/ \tau)^\beta$)}
\label{fig4}
\end{figure}

We shall now discuss the characteristic metastable behaviour associated with the RSG and MG state. The metastable response of SG and RSG systems continues to remain a subject of active interest \cite{23}. The ZFC state of these systems show strong relaxation in magnetization, while the field cooled state does not \cite{12,13}. These behaviors are exemplified in Fig. 4 (a), which shows M versus time (t) plots for Au$_{82}$Fe$_{18}$ at T=14K and H=100 Oe in the ZFC and FCC state. The M(t) data in the ZFC state can be filled with the equation M(t)= -1 +2t$^\gamma$,where $\gamma$ indicates the extent of relaxation. Higher value of $\gamma$ means a greater degree of relaxation in the same t interval. This equation has earlier been shown to apply to the relaxation of ferromagnetic dots which interact through long range dipolar interaction \cite{24}. The obtained value of exponent $\gamma$ in the present case is 0.004. Since the ZFC state in MG is an equilibrium state, no relaxation of M is observed there. On the other hand, entrance in to the MG state along the FCC path introduces distinct glass-like relaxation and the divergence of the relaxation time with lowering in T \cite{2,7,9}.  To contrast such metastabilty with that observed in the RSG state, we present in Fig. 4(b) M vs time plot obtained for Ce(Fe$_{0.96}$Ru$_{0.04}$)$_2$ at T=18K and H=20 kOe in the ZFC state and FCC state. The M(t) data in the FCC state can be filled with the Kohlrausch-Williams-Watt (KWW) stretched exponential function M(t) $\propto$ exp[-(t$/ \tau)^\beta$, where $\tau$ is the charactteristic relaxation time and $\beta$ is a shape parameter \cite{2}. The obtained value of exponent $\beta$ here is 0.65. 

The metastable nature of the FCC state in systems showing MG behaviour can be supported further by showing that this state is susceptible to any energy fluctuation introduced by a T or H cycling \cite{2,7,9}. Similar extensive T cycling in the FCC state of the present Au$_{82}$Fe$_{18}$ sample failed to reveal any such signature of metastablity \cite{25}.    

Summarizing the above experimental results, we identify four distinct experimental features in bulk magnetization measurements, which can be used to distinguish a magnetic-glass from a reentrant spin-glass:

(i)	MG arises out of the kinetic arrest of a first order FM-AFM phase transition. This first order transition will give rise to a distinct thermal hystersis between the FCC and FCW magnetization. No such thermal hysteresis is expected in the case of FM (or AFM)-RSG transition, since this is considered to be a second order phase transition \cite{14} or a gradual phase transformation \cite{15}.

(ii)The TMI decreases with the increase in applied H in RSG systems. This is just the opposite in MG, where TMI appears only above an critical applied H (the value of which will depend on the system under consideration) and increases with the increase in H.

(iii) A newly introduced experimental protocol 'cooling and heating in unequal field (CHUF)' reveals distinct features in the T dependence of magnetization in MG, which depend on the sign of inequality between the fields applied during cooling and heating. No such features are expected for a RSG system.

(iv) ZFC state of RSG shows distinct relaxation in magnetization, while the FC state does not. The behaviour observed in the MG systems is just the opposite.

In conclusion the newly observed magnetic-glass phenomenon in different magnetic systems is distinctly different from the well known spin-glass and re-entrant spin-glass phenomena. While the existence of a new magnetic-glass systems will finally be established through microscopic studies like magnetic imaging, the experimental criteria described in this work can definitely be used for regular identification of a magnetic-glass. 

The authors thank P. Chaddah for useful discussion.


\begin{thebibliography}{99}
\bibitem{1}M. A. Manekar, S. Chaudhary, M. K. Chattopadhyay, K. J. Singh, S. B. Roy and P. Chaddah, Phys, Rev. {\bf B64} 104416 (2001).  
\bibitem{2} M. K. Chattopadhyay, S. B. Roy and P. Chaddah Phys. Rev. {\bf B72} 180401R (2005).
\bibitem{3} K. Kumar A. K. Pramanik, A. Banerjee, P. Chaddah, S. B. Roy, S. Park, C. L. Zhang, and S.-W. Cheong, Phys.Rev. {\bf B 73},184435 (2006).
\bibitem{4} W. Wu, C. Israel, N. Hur, P. Soonyong, S.-W. Cheong, and A. De Lozane, Nat. Mater.{\bf 5}, 881 (2006).
\bibitem{5} A. Banerjee, K. Mukherjee, Kranti Kumar, and P. Chaddah, J. Phys. :Condens. Mater.{\bf 18}, L605 (2006); A. Banerjee, K. Mukherjee, Kranti Kumar, and P. Chaddah,,Phys. Rev. {\bf B74}, 224445 (2006); Z W Ouyang,H. Nojiri and S. Yoshii, Phys. Rev. {\bf B78} 104404 (2008)
\bibitem{6} K. Sengupta and E. V. Sampathkumaran, Phys. Rev. {\bf B73}, 020406R (2006); P. Kushwaha, R. Rawat and P. Chaddah, J. Phys.:Condens. Matr. {\bf 20}, 022204 (2008).
\bibitem{7} S. B. Roy, M. K. Chattopadhyay, P. Chaddah,J. D. Moore, G. K. Perkins, L. F. Cohen, K. A. Gschneidner, Jr. and V. K. Pecharsky, Phys. Rev. {\bf B74}, 012403 (2006).
\bibitem{8} P G Debenedetti and F H Stillinger, Natur {\bf 410}, 259 (2001).
\bibitem{9} V K Sharma, M. K. Chattopadhyay and S B Roy, Phys. Rev. {\bf B76}, 140401(2007). 
\bibitem{10} S. B. Roy, M. K. Chattopadhyay, A. Banerjee, P. Chaddah, J. D.Moore, G. K. Perkins, L. F. Cohen, K. A. Gschneidner Jr., and V.
K. Pecharsky, Phys Rev.{\bf B75}, 184410 (2007); M. K. Chattopadhyay, S. B. Roy, K. Morrison, J. D.Moore, G. K. Perkins, L. F. Cohen, K. A. Gschneidner Jr., and V.K. Pecharsky,  Euro.Phys. Lett. {\bf 83} 57006 (2008).
 \bibitem{11}P. Chaddah, Kranti Kumar and A Banerjee, Phys. Rev. {\bf B67}, 100402(R) (2008). 
\bibitem{12} K. H. Binder and A. P. Young, Rev. Mod. Phys. {\bf 58} 801 (1986).
\bibitem{13} J A Mydosh, Spin Glasses (Taylor and Francis, 1992).
\bibitem{14} M. Gabay and G. Toulouse, Phys. Rev. Lett. {\bf 47}, 201 (1981).
\bibitem{15} S. Nidera and F. Matsubara, 2007, Phys. Rev. {\bf B75} 144413 (2007).
\bibitem{16} S B Roy, G. K. Perkins, M. K. Chattopadhyay, A. K. Nigam, K. J. S. Sokhey, P. Chaddah, A. D. Caplin and L. F. Cohen, Phys. Rev. Lett. {\bf 92}, 147203 (2004). 
\bibitem{17} B. R. Coles, B. V. B. Sarkissian and R. H. Taylor, Philos. Mag. {\bf B37}, 489 (1978); B. V. B. Sarkissian, J. Phys. {\bf F 11}, 2191 (1981);I. A. Campbell, D. Arvanitis and A. Fert, Phys. Rev. Lett. {\bf 51} 57 (1983). 
\bibitem{18} A. K. Gangopadhyay, S. B. Roy and A. K. Majumdar, Phys. Rev. {\bf B33}, 5010 (1986).
\bibitem{19} M H Bhat,V Molinero, E Soignard E, V C Solomon, S Sastry and J L Yarger, Nature {\bf 448},787 (2007). 
\bibitem{20} Y Imry and M Wortis, Phys. Rev. {\bf B19}, 3580 (1979).
\bibitem{21} P Chaddah, A. Banerjee and S. B. Roy Cond-Mat arXiv:0601095 (2006).
\bibitem{22} A. Banerjee, K. Kumar and P. Chaddah, J. Phys.: Condens. Matter {\bf 21} 026002(2009);S Dash, A. Banerjee and P. Chaddah, Solid. St. Commun.{\bf 148} 336 (2008)
\bibitem{23} K. Hioki and K. Motoya, J. Phys. Soc. Japan, {\bf 74}, 1830 (2005) (and references within).
\bibitem{24} L. C. Sampaio,R. Hyndman,F. S. de Menezes, J. P. Jamet, P. Meyer, J. Gierak,C. Chappert,
V. Mathet and J. Ferre , Phys. Rev. {\bf B64}, 18440 (2001).
\bibitem{25} S B Roy and M K Chattopadhyay (unpublished results).

\end{thebibliography}
\end{document}